\documentclass[12pt, a4paper]{article}
\usepackage[font=footnotesize]{caption}
\usepackage[utf8]{inputenc}
\usepackage{graphicx}
\usepackage{authblk}
\usepackage{booktabs}
\usepackage{verbatim}
\usepackage{url}
\usepackage{caption}
\usepackage{subcaption}
\usepackage{indentfirst}
\usepackage{url}
\usepackage{amssymb}
\usepackage{amsmath}
\title{Detecting Groups in Directed and Non-Directed Bipartite Networks}

\author{Alexandre Benatti$^1$ and Luciano da F. Costa$^2$}

\affil{
$^1$Institute of Mathematics and Statistics - DCC \\
University of S\~ao Paulo \\
Rua do Mat\~ao, 1010, S\~ao Paulo, SP 05508-090 Brazil 
\\ \vspace{0.5cm}
$^2$S\~ao Carlos Institute of Physics - DFCM \\
University of S\~ao Paulo \\
Av. Trabalhador S\~ao-Carlense, 400, S\~ao Carlos, SP 13566-590 Brazil
}

\date{5th Dec. 2023}

\begin{document}

\maketitle

\begin{abstract}
Bipartite networks provide an effective resource for representing, characterizing, and modeling several abstract and real-world systems and structures involving binary relations, which include food webs, social interactions, and customer-product relationships. Of particular interest is the problem of, given a specific bipartite network, to identify possible respective groups or clusters characterized by similar interconnecting patterns. The present work approaches this issue by extending and complementing a previously described coincidence similarity methodology (Bioarxiv, doi.org/10.1101/2022.07.16.500294) in several manners, including the consideration of direct and non-directed bipartite networks, the characterization of groups in those networks, as well as considering synthetic bipartite networks presenting groups as a resource for studying the performance of the described methodology. Several interesting results are described and discussed, including the corroboration of the potential of the coincidence similarity methodology for achieving enhanced separation between the groups in bipartite networks.
\end{abstract}

\section{Introduction}\label{sec:introduction}

Bipartite networks can be understood as a specific type of complex network intrinsically characterized by binary relationships (i.e.~relationships between pairs of elements). Examples include but are by no means limited to food webs (e.g.~\cite{saavedra2009simple,bezerra2009pollination,eklof2013dimensionality,pavlopoulos2018bipartite}), customer preferences (e.g.~\cite{zaccaria2014taxonomy,dahui2006bipartite,shang2010empirical}), and social interactions (e.g.~\cite{lambiotte2005uncovering,bramoulle2007anti,burgos2008two}).

Given a particular bipartite network, it is often the case that it needs to be analyzed in order to identify repeating patterns (or groups/clusters) of binary interactions, such as between two or more types of consumers and respective resources. Several interesting related approaches have been reported, including~\cite{lind2005cycles,barber2007modularity,liu2010efficient,guimera2007module,dormann2014method,larremore2014efficiently,beckett2016improved}, which often rely on the estimation of the modularity of projections (e.g.~\cite{calderer2021community,barber2007modularity,larremore2014efficiently}) of the original bipartite network.

More recently~\cite{costa_biorXiv}, an approach involving the representation of food webs in terms of respective coincidence similarity networks was reported which showed promising potential for enhancing the identification of clusters in bipartite networks. This methodology is intrinsically based on the quantification of the similarity between the connections from the first to the second column in a bipartite network, which is estimated by using the coincidence similarity index~\cite{da2021further, costa2022on,CostaCCompl,costa2023multiset}, which has interesting properties including strict comparison, robustness to outliers, and intrinsic normalization.

Though the potential of the coincidence similarity approach to bipartite network analysis was preliminaryillustrated in~\cite{costa_biorXiv}, that work was restricted to some real-world bipartite networks. The present work extends that previous work in several manners. First, it describes direct and non-directed bipartite networks, emphasizing the possibility of having direct and reverse representations in the latter case. Second, it discusses the important issue of having groups (or clusters) in a bipartite network, and how this type of network can be systematically modeled to obtain a virtually unlimited number of synthetic bipartite networks presenting modular structure. This interesting resource is then considered as the basis for investigating the potential of the proposed similarity methodology respectively to the identification of clusters in synthetic non-directed, weighted bipartite networks containing varying numbers of groups, number of elements per group, as well as varying levels of irregularity in the interconnections within each group. Several interesting results are described and discussed, including the identification of the effectiveness of the coincidence similarity approach in identifying groups of nodes even in cases where the interconnections within each group are substantially irregular.

This work starts by presenting the basic concepts and methods, which include bipartite networks, clusters in these types of networks, and the coincidence similarity index. Several experimental results are then described and discussed, which is then followed by concluding remarks and prospects for future related works.

\section{Concepts and Methods}\label{sec:methods}

This section presents the main concepts and methods adopted in the study of bipartite networks represented as similarity networks, which include bipartite networks, clusters in this type of networks, the coincidence similarity index, as well as the procedure adopted for generating synthetic bipartite networks containing pre-specified groups.

\subsection{Bipartite Networks}

\begin{figure}[h!]
  \centering
     \includegraphics[width=.3 \textwidth]{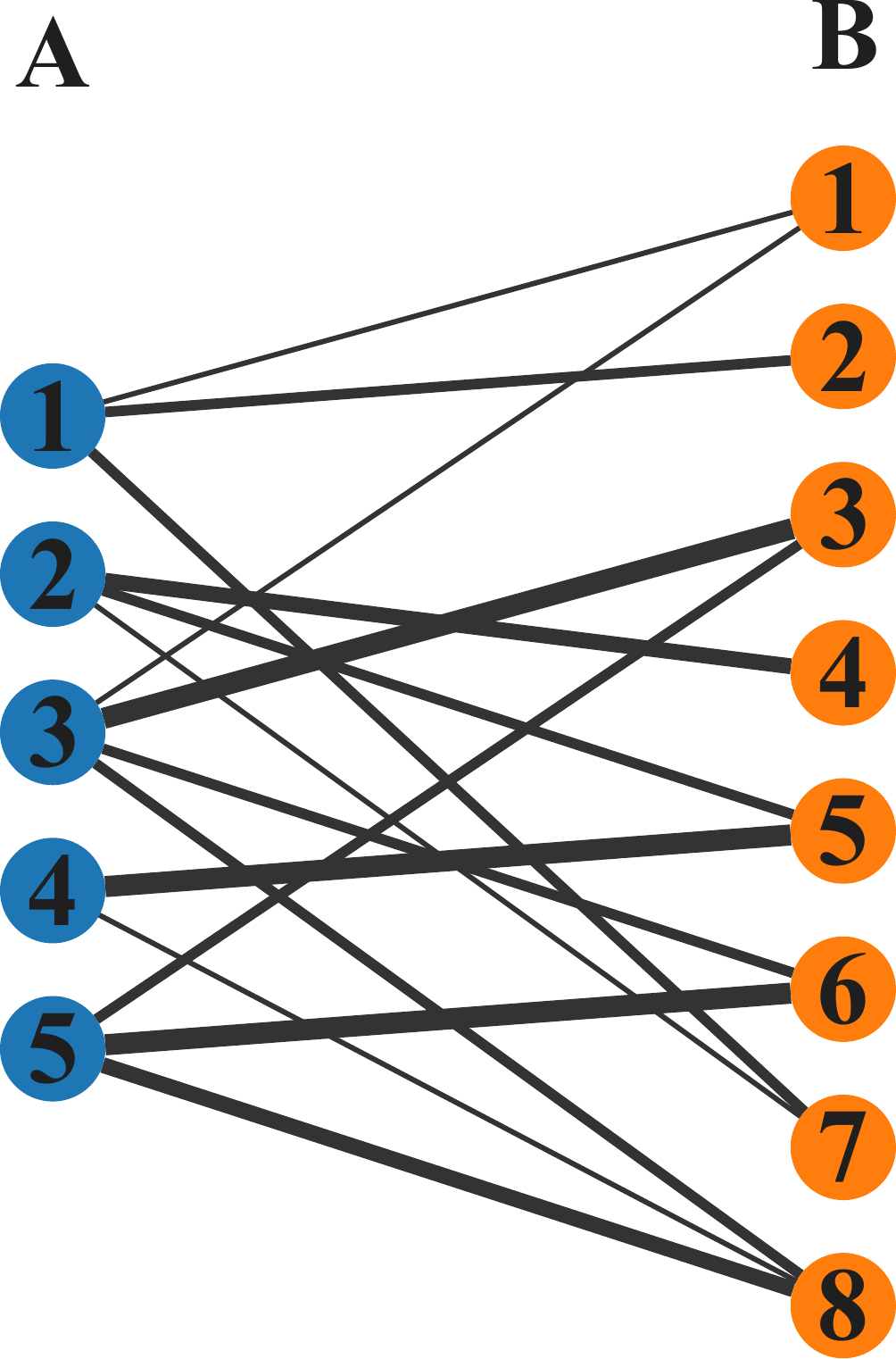}
   \caption{Example of a non-directed, weighted bipartite network involving 5 nodes of type $A$ and 8 nodes of type $B$. Interconnections are only possible between nodes of distinct types. The width of each interconnections indicate the respective weight. Except for incorporating two type of nodes, this structure can actually be understood as a classic non-directed, weighted graph with 13 nodes.}\label{fig:bipartite}
\end{figure}

Bipartite networks can be understood as a particular type of graph involving two types of nodes -- henceforth identified as $A$ (left) and $B$ (right) -- with interconnections allowed only between nodes of different types. Other than that, a bipartite network is henceforth understood to be a classic graph, which can be represented in terms of adjacency or weight matrices. Figure~\ref{fig:bipartite} illustrates a simple non-directed, weighted bipartite network involving 5 nodes of type $A$ and 8 nodes of type $B$. Observe that the two types of nodes are typically organized into two respective columns.

Table~\ref{tab:bipartite} presents the $13 \times 13$ weight matrix indicating the interconnections between the nodes in the bipartite shown in Figure~\ref{fig:bipartite}.

\begin{table}[h!]
\centering
\caption{Interconnection (weight) matrix respective to the non-directed, weighted bipartite network shown in Fig.~\ref{fig:bipartite}. The values of each entry correspond to the respective weight in the bipartite network. Observe that this matrix is necessarily symmetric.}\label{tab:bipartite}
\begin{tabular}{c|ccccccccccccc|}
\cline{2-14}
\textbf{}                         & \textbf{A1} & \textbf{A2} & \textbf{A3} & \textbf{A4} & \textbf{A5} & \textbf{B1} & \textbf{B2} & \textbf{B3} & \textbf{B4} & \textbf{B5} & \textbf{B6} & \textbf{B7} & \textbf{B8} \\ \hline
\multicolumn{1}{|c|}{\textbf{A1}} & 0           & 0           & 0           & 0           & 0           & 1           & 2           & 0           & 0           & 0           & 0           & 2           & 0           \\
\multicolumn{1}{|c|}{\textbf{A2}} & 0           & 0           & 0           & 0           & 0           & 0           & 0           & 0           & 3           & 2           & 0           & 1           & 0           \\
\multicolumn{1}{|c|}{\textbf{A3}} & 0           & 0           & 0           & 0           & 0           & 1           & 0           & 4           & 0           & 0           & 2           & 0           & 2           \\
\multicolumn{1}{|c|}{\textbf{A4}} & 0           & 0           & 0           & 0           & 0           & 0           & 0           & 0           & 0           & 4           & 0           & 0           & 1           \\
\multicolumn{1}{|c|}{\textbf{A5}} & 0           & 0           & 0           & 0           & 0           & 0           & 0           & 2           & 0           & 0           & 3           & 0           & 3           \\
\multicolumn{1}{|c|}{\textbf{B1}} & 1           & 0           & 1           & 0           & 0           & 0           & 0           & 0           & 0           & 0           & 0           & 0           & 0           \\
\multicolumn{1}{|c|}{\textbf{B2}} & 2           & 0           & 0           & 0           & 0           & 0           & 0           & 0           & 0           & 0           & 0           & 0           & 0           \\
\multicolumn{1}{|c|}{\textbf{B3}} & 0           & 0           & 4           & 0           & 2           & 0           & 0           & 0           & 0           & 0           & 0           & 0           & 0           \\
\multicolumn{1}{|c|}{\textbf{B4}} & 0           & 3           & 0           & 0           & 0           & 0           & 0           & 0           & 0           & 0           & 0           & 0           & 0           \\
\multicolumn{1}{|c|}{\textbf{B5}} & 0           & 2           & 0           & 4           & 0           & 0           & 0           & 0           & 0           & 0           & 0           & 0           & 0           \\
\multicolumn{1}{|c|}{\textbf{B6}} & 0           & 0           & 2           & 0           & 3           & 0           & 0           & 0           & 0           & 0           & 0           & 0           & 0           \\
\multicolumn{1}{|c|}{\textbf{B7}} & 2           & 1           & 0           & 0           & 0           & 0           & 0           & 0           & 0           & 0           & 0           & 0           & 0           \\
\multicolumn{1}{|c|}{\textbf{B8}} & 0           & 0           & 2           & 1           & 3           & 0           & 0           & 0           & 0           & 0           & 0           & 0           & 0           \\ \hline
\end{tabular}
\end{table}

\begin{figure}[h!]
  \centering
     \includegraphics[width=.3 \textwidth]{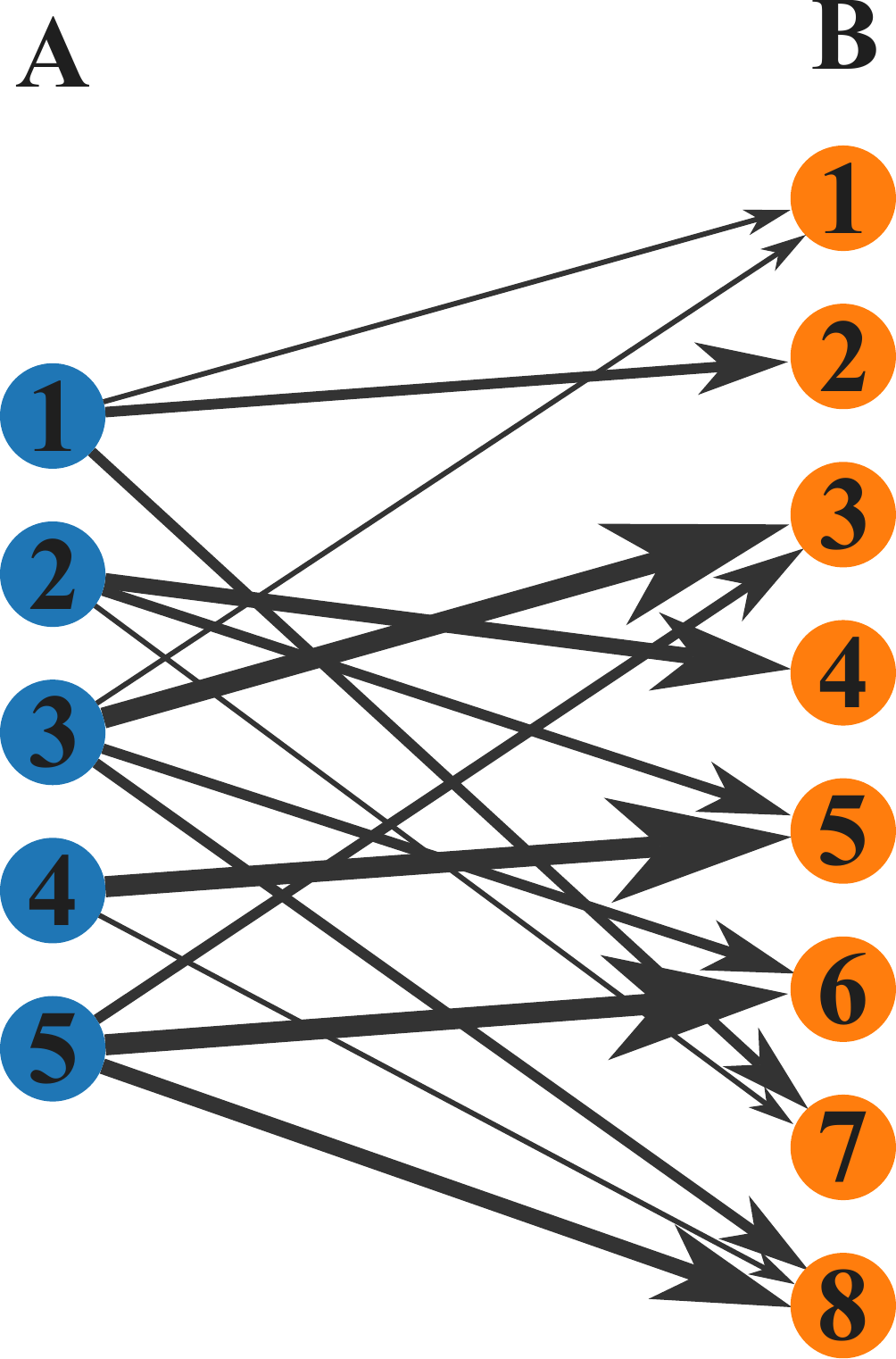}  \hspace{2cm}
     \includegraphics[width=.3 \textwidth]{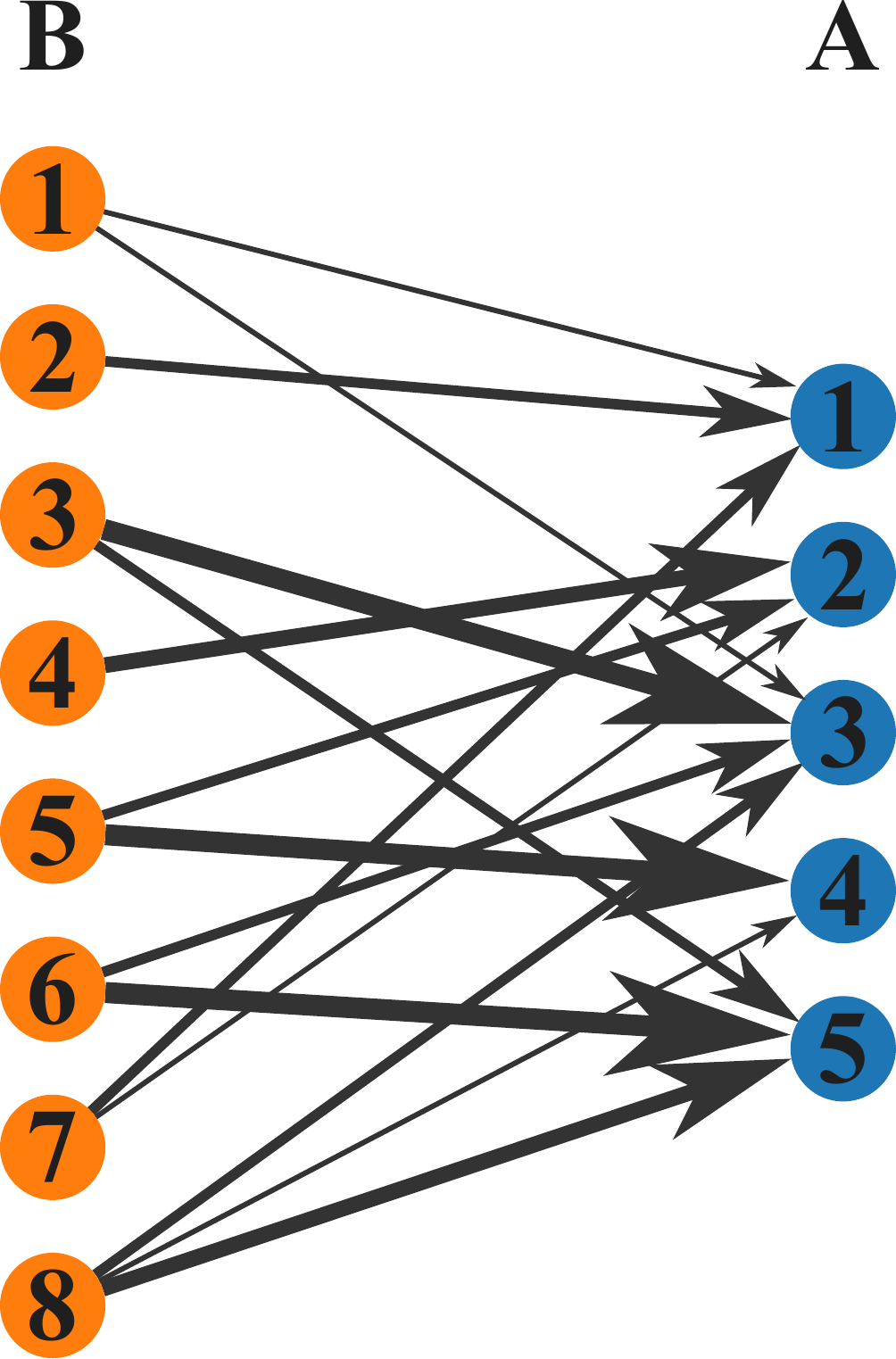} \\
     (a) \hspace{6cm} (b)
   \caption{The non-directed network in Fig.~\ref{fig:bipartite} can be understood in terms of directed bipartite networks corresponding to \emph{direct} (a) and \emph{reverse} (b) representations. The latter type of representation is only applicable in the case of non-directed bipartite networks.}\label{fig:bipartite_D}
\end{figure}

\begin{table}[h!]
\centering
\caption{Interconnection matrix respective to the \emph{direct} bipartite network obtained fro the structure shown in Fig.~\ref{fig:bipartite_D}(a). The rows correspond to the nodes of type $A$, and the matrix entries indicate the weight of the interconnection between those nodes and the nodes of type $B$, represented as columns. In the case of directed bipartite networks, the connections are assumed to proceed from nodes in the rows to nodes in the columns.}\label{tab:bipartiteA}
\begin{tabular}{c|cccccccc|}
\cline{2-9}
\textbf{}                          & \textbf{B1} & \textbf{B2} & \textbf{B3} & \textbf{B4} & \textbf{B5} & \textbf{B6} & \textbf{B7} & \textbf{B8} \\ \hline
\multicolumn{1}{|c|}{\textbf{A1}} & 1            & 2            & 0            & 0            & 0            & 0            & 2            & 0            \\
\multicolumn{1}{|c|}{\textbf{A2}} & 0            & 0            & 0            & 3            & 2            & 0            & 1            & 0            \\
\multicolumn{1}{|c|}{\textbf{A3}} & 1            & 0            & 4            & 0            & 0            & 2            & 0            & 2            \\
\multicolumn{1}{|c|}{\textbf{A4}} & 0            & 0            & 0            & 0            & 4            & 0            & 0            & 1            \\
\multicolumn{1}{|c|}{\textbf{A5}} & 0            & 0            & 2            & 0            & 0            & 4            & 0            & 3            \\ \hline
\end{tabular}

\end{table}

Though the above mentioned representation of bipartite networks has been frequently adopted, there are other complementary manners in which to understand a bipartite network. For instance, bipartite networks can be projected into any of the two types of nodes, $A$ or $B$ (e.g.~\cite{calderer2021community,barber2007modularity,larremore2014efficiently}), yielding a respective unipartite network.  In the case the projection is to be performed within the nodes of type $A$, the resulting network will contain these nodes, the connection between any pair of them being a link typically weighted by the number of shared nodes of type $B$.  

\begin{figure}[h!]
  \centering
     \includegraphics[width=.4 \textwidth]{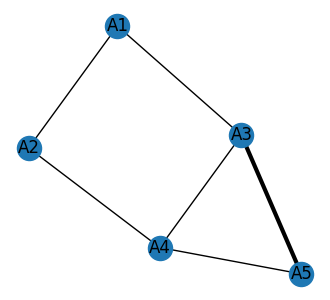}  \hspace{1cm}
     \includegraphics[width=.4 \textwidth]{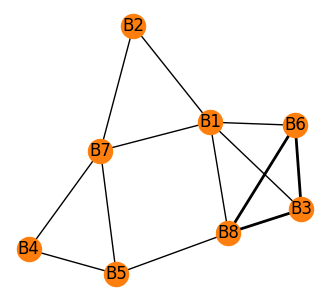}  \\
     (a) \hspace{6cm} (b)
   \caption{The projections of the bipartite network shown in Fig.~\ref{fig:bipartite} into the $A$  and $B$ nodes, shown respectively in (a) and (b).  These results do not take into account the weights of the original bipartite network.}\label{fig:Proj}
\end{figure}

Figure~\ref{fig:Proj}(a) and (b) illustrate the projections of the bipartite network in Figure~\ref{fig:bipartite} into the sets of nodes of type $A$ and $B$, respectively.  The weights of the original bipartite network are not taken into account in this example.

Yet another approach to study bipartite networks has been described in ~\cite{costa_biorXiv}, where a table is adopted for specifying the interconnections between the nodes of type $A$, organized as rows of the table, with the nodes of type $B$, represented as table columns. The columns can be understood as being analogous to the \emph{features} or \emph{properties} of the row entries. Table~\ref{tab:bipartite} illustrates the obtained respectively from the bipartite network shown in Figure~\ref{fig:bipartite}.  As a matter of fact, the table representation of a bipartite network corresponds to the way in which this type of structure is often initially provided.

It is important to observe that the interconnections between the nodes of type $A$ and $B$ can be \emph{directed} (connections extending from left to right) or \emph{non-directed}, indicating symmetric and asymmetric binary relationships. In the latter case, it is possible to consider not only the \emph{direct} connections from $A$ to $B$, but also vice-versa. 

The directions of the links in a directed bipartite network are henceforth understood to indicate that the element from which the links originate is characterized, or has the outgoing links as respective properties. Therefore, the grouping of nodes should take place only respectively to the nodes from which the links originate (of one type) while considering only the properties specified in the respective destination nodes (of other type).  

Figure~\ref{fig:bipartite_D} depicts both the \emph{direct} and \emph{reverse} forms of the bipartite network in Figure~\ref{fig:bipartite}.  Observe that these two forms actually correspond to respective directed bipartite networks.  The interconnections respective to the directed bipartite networks shown in Figure~\ref{fig:bipartite_D}(a) and (b) are presented in Tables~\ref{tab:bipartite} and~\ref{tab:bipartiteB}.

\begin{table}[h!]
\centering
\caption{Interconnection matrix respective to the \emph{reverse} bipartite network obtained from the network in Fig.~\ref{fig:bipartite_D}((b). The rows correspond to the nodes of type $A$, and the matrix entries indicate the weight of the interconnection between those nodes and the nodes of type $B$, represented as columns.}\label{tab:bipartiteB}
\begin{tabular}{c|ccccc|}
\cline{2-6}
\textbf{}                          & \textbf{A1} & \textbf{A2} & \textbf{A3} & \textbf{A4} & \textbf{A5} \\ \hline
\multicolumn{1}{|c|}{\textbf{B1}} & 1            & 0            & 1            & 0            & 0   \\
\multicolumn{1}{|c|}{\textbf{B2}} & 2            & 0            & 0            & 0            & 0   \\
\multicolumn{1}{|c|}{\textbf{B3}} & 0            & 0            & 4            & 0            & 2   \\
\multicolumn{1}{|c|}{\textbf{B4}} & 0            & 3            & 0            & 0            & 0   \\
\multicolumn{1}{|c|}{\textbf{B5}} & 0            & 2            & 0            & 4            & 0   \\
\multicolumn{1}{|c|}{\textbf{B6}} & 0            & 0            & 2            & 0            & 4   \\
\multicolumn{1}{|c|}{\textbf{B7}} & 2            & 1            & 0            & 0            & 0   \\
\multicolumn{1}{|c|}{\textbf{B8}} & 0            & 0            & 2            & 1            & 3    \\ \hline
\end{tabular}

\end{table}

It should be kept in mind that, in the case of non-directed bipartite networks, any of the two respective interconnection matrices (i.e.~direct or reversed) will be enough to completely specify the respective network. However, the interpretation of the two tables is typically distinct, being related to the respectively chosen direction of the links.

\begin{figure}[h!]
  \centering
     \includegraphics[width=.5 \textwidth]{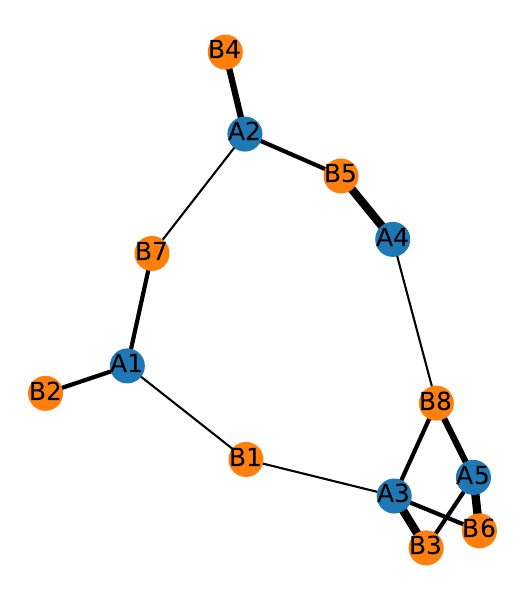}
   \caption{The bipartite network in Fig.~\ref{fig:bipartite} with nodes visualized irrespectively to their types. It is interesting to observe the markedly distinct perception of the same bipartite network motivated by these two alternative representations.}\label{fig:bipartite_net}
\end{figure}

Another interesting manner to visualize bipartite networks is as shown in Figure~\ref{fig:bipartite_net}, where the nodes of type $A$ and $B$ have been distributed irrespectively to their types by using a network visualization method (e.g.~Fruchterman-Reingold~\cite{fruchterman1991graph}). This type of visualization is henceforth referred to as the \emph{unconstrained} version of a bipartite network.

Because the case of directed bipartite networks is in this work intrinsically incorporated in the described analysis of non-directed bipartite networks, attention is henceforth focused on the direct and reverse representations obtained from non-directed bipartite networks.

\subsection{Clusters in Bipartite Networks}

\begin{figure}[h!]
  \centering
     \includegraphics[width=.7 \textwidth]{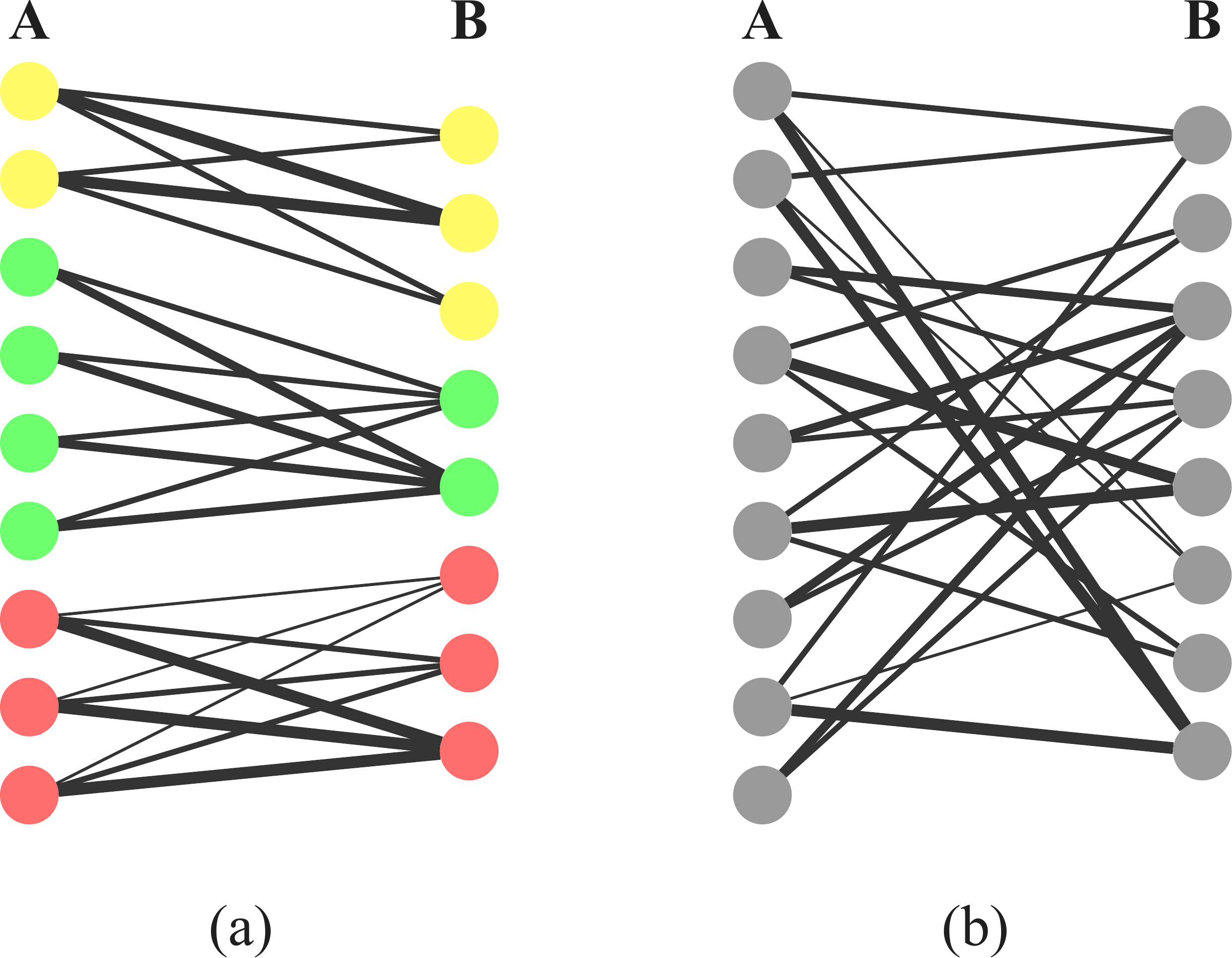}
   \caption{A non-directed, weighted bipartite network presenting completely separated (no overlap of interconnections) clusters of nodes (a), and the same network represented with nodes in arbitrary order. Typically, a network to be studied is given as in (b), and cluster identification methods need to be applied in order to obtain representations with identified clusters as that illustrated in (a).}\label{fig:bipartite2}
\end{figure}

Though bipartite networks can have any structure, a situation of particular interest concerns the presence of groups or clusters among the nodes of the respective bipartite structure. Figure~\ref{fig:bipartite2} illustrates a non-directed bipartite network containing 3 groups of nodes of type $A$. Observe that the clusters are characterized by groupings of both types of nodes.

In the present work, we shall understand as a \emph{group} (or cluster) of nodes of type $A$ in a given bipartite network the set of nodes which have identical or similar patterns of interconnection (within each respective group) with the nodes of type $B$. In the case of the above example, we have three groups of nodes of type $A$, each group presenting identical interconnections with nodes of type $B$.  The type structure illustrated in Figure~\ref{fig:bipartite2}(a), presenting completely separated groups, can be understood as the \emph{reference model} of bipartite networks presenting groups as considered in the present work.  

The identification of clusters in a representation such as that illustrated in Figure~\ref{fig:bipartite2}(a) is immediate as a direct consequence of the organization of the groups as subsequent sets of nodes. More generally, networks to be analyzed respectively to clustering are typically provided in a generic manner, as illustrated in Figure~\ref{fig:bipartite2}(b) respectively to the very same bipartite network shown in Figure~\ref{fig:bipartite2}(a). Now, the identification of possible groups becomes nearly impossible to be performed through human visual inspection, motivating the development of effective computational approaches capable of effectively finding possible groups in a given bipartite network.

\begin{figure}[h!]
  \centering
     \includegraphics[width=.25 \textwidth]{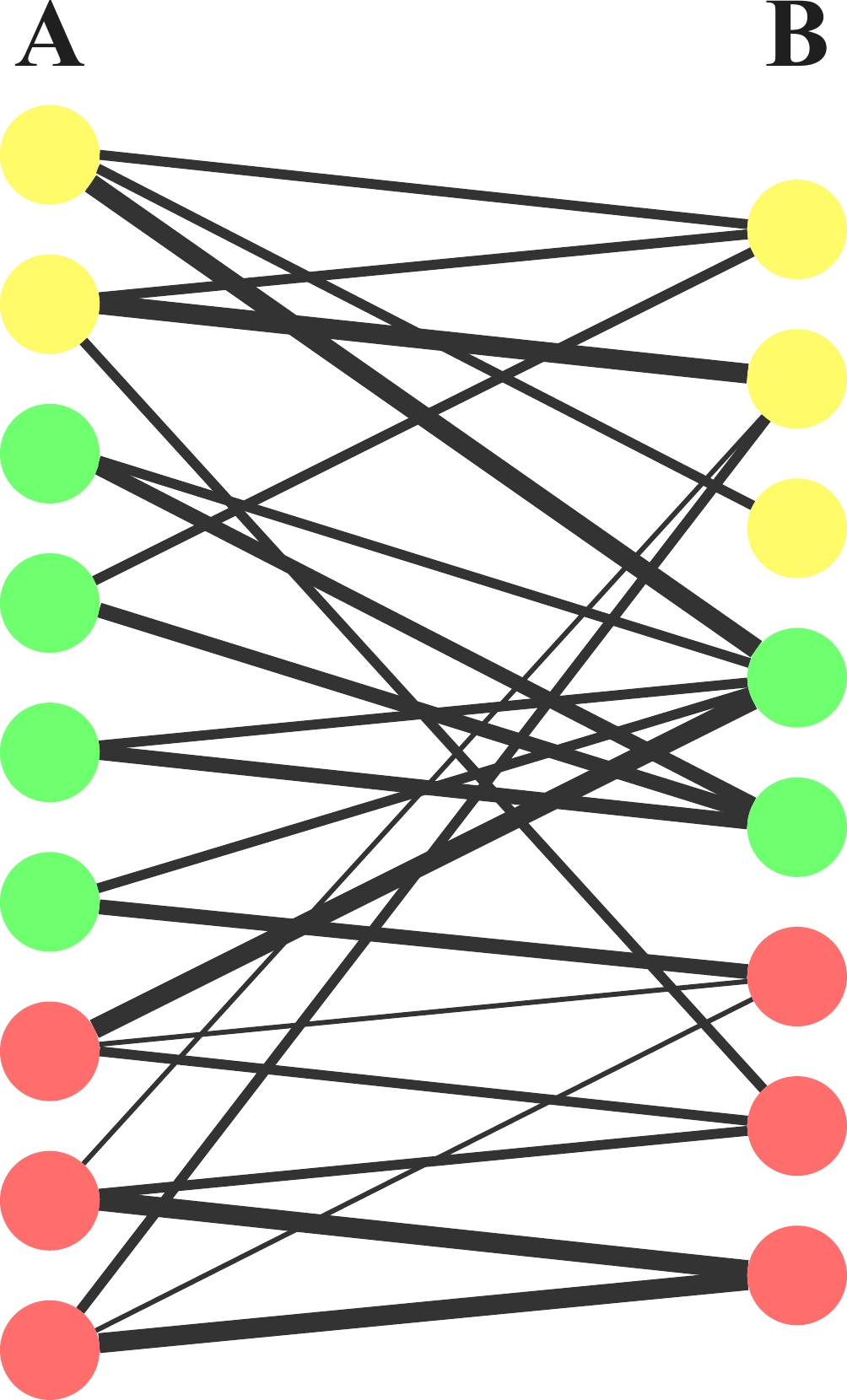}
   \caption{A non-directed, weighted bipartite network containing three groups of nodes with some overlaps between the interconnections. The identification of groups in the presence of overlap is intrinsically more challenging (or even impossible in some cases) than when the groups are completely separated as in Fig.~\ref{fig:bipartite2}.}\label{fig:bipartite3}
\end{figure}

The problem of identifying groups in bipartite networks is often substantially compounded by the presence of overlapping interconnections between distinct groups, as illustrated in Figure~\ref{fig:bipartite3}.  Because real-world bipartite networks almost invariably present some level of overlap, effective methods for identifying groups in this type of networks become necessary, which provides the main motivation for the present work and described methodology based on the coincidence similarity index.

\subsection{The Coincidence Similarity Index}

Informally speaking, a group (or cluster) in a set of $N$ data elements, each characterized by $M$ respective features (or properties), can be understood as a subset of that dataset presenting features that are more similar within that set than with the remainder of the elements. Frequently the similarity between a pair of data elements is quantified by resourcing to a respective \emph{distance} (e.g.~Euclidean, Minkowski, Manhattan), implicitly assuming that the more distant a pair of data elements are less similar.

The approaches to cluster identification employing similarity have often relied on the \emph{cosine similarity} (e.g.~\cite{xia2015learning,luo2018cosine}) and cross-correlations (e.g.~\cite{costa2022on}). More recently, the \emph{coincidence similarity index} was described~\cite{da2021further,costa2022on,CostaCCompl} as a means to quantify the similarity between a pair of non-zero vectors that is more strict than the Jaccard similarity index (e.g.~\cite{Jaccard1,Jac:wiki,Loet}), while presenting some additional interesting characteristics, including robustness to outliers and intrinsic normalization. Basically, the coincidence similarity index corresponds to the product of the respective Jaccard and interiority (or overlap, e.g.~\cite{vijaymeena}) indices.

The approach to the identification of groups in bipartite networks considered in the present work adopts the coincidence similarity index as the means for transforming a given bipartite network into a respective coincidence similarity network~\cite{CostaCCompl,benatti2022neuromorphic}, as described in the following.

Given two non-empty sets $A$ and $B$, each with $N$ elements, the \emph{Jaccard similarity index} (e.g.~\cite{Jaccard1,Jac:wiki,Loet}) between those two sets can be expressed as:
\begin{align}
    \mathcal{J}(A,B) = \frac{A \cap B}{A \cup B}
\end{align}

with $0 \leq \mathcal{J}(A,B) \leq 1$.

The \emph{interiority} (or \emph{overlap}~\cite{vijaymeena}) index between the two above sets can be calculated as:
\begin{align}
    \mathcal{I}(A,B) = \frac{A \cap B}{\min \left\{|A|, |B| \right\} }
\end{align}

with $0 \leq \mathcal{I}(A,B) \leq 1$.

The \emph{coincidence similarity index} has been described~\cite{da2021further,costa2022on,CostaCCompl} as an enhanced index to address the fact that the Jaccard index cannot take into account the interiority between the two compared sets. More specifically, we have that the coincidence similarity index between two non-empty sets $A$ and $B$ has been defined as:
\begin{align}
    \mathcal{C}(A,B) = \mathcal{J}(A,B) \, \mathcal{I}(A,B)
\end{align}

also with $0 \leq \mathcal{C}(A,B) \leq 1$.

By generalizing sets as multisets (or vectors) $\vec{x}$ and $\vec{y}$ (e.g.~\cite{da2022multisets}), which are here assumed to be non-zero and contain non-negative real-aued elements, the Jaccard, interiority, and coincidence similarity indices can be re-written as follows:
\begin{align}  \label{eq:J}
    \mathcal{J}(\vec{x},\vec{y}) = \frac{\sum_{i=1}^N \min \left\{x_i, y_i \right\} }{\sum_{i=1}^N \max \left\{x_i, y_i \right\}}
\end{align}

\begin{align}  \label{eq:I}
    \mathcal{I}(\vec{x},\vec{y}) = \frac{\sum_{i=1}^N \min \left\{x_i, y_i \right\}}{\min \left\{\sum_{i=1}^N x_i, \sum_{i=1}^N y_i \right\} }
\end{align}

\begin{align}   \label{eq:C}
    \mathcal{C}(\vec{x},\vec{y}) = \mathcal{J}(\vec{x},\vec{y}) \, \mathcal{I}(\vec{x},\vec{y})
\end{align}

All the three indices above have values within the interval $[0,1]$.  In case the vector elements can take negative values, adapted forms of the above equation can be obtained by using the more general form of multiset intersection described in~\cite{da2023mulsetions}.

As a simple example of numeric calculation of the coincidence similarity index between two vectors, let us consider $\vec{x} = \left[2, 0, 1, 3.5 \right]$ and $\vec{y}= \left[1.2, 3, 2, 0 \right]$.  The Jaccard similarity index between these vectors can be obtained by using  Equation~\ref{eq:J}, yielding:
\begin{align}
  &\mathcal{J}(\vec{x},\vec{y}) = \frac
  {\min \left\{2,1.2 \right\} + \min \left\{0,3 \right\} +
  \min \left\{1,2 \right\} + \min \left\{3.5,0 \right\}}
  {\max \left\{2,1.2 \right\} + \max \left\{0,3 \right\} +
  \max \left\{1,2 \right\} + \max \left\{3.5,0 \right\}} =
  \nonumber \\
  &= \frac{1.2+0+1+0}{2+3+2+3.5} = \frac{2.2}{10.5}  \nonumber
\end{align}

The respective interiority index can be calculated by using Equation~\ref{eq:I} as follows:
\begin{align}
  &\mathcal{I}(\vec{x},\vec{y}) = \frac
  {\min \left\{2,1.2 \right\} + \min \left\{0,3 \right\} +
  \min \left\{1,2 \right\} + \min \left\{3.5,0 \right\}} 
  {\min \left\{6.5, 6.2 \right\}} =\frac{2.2}{6.2}    \nonumber
\end{align}

Now, the coincidence index between the vectors $\vec{x}$ and $\vec{y}$ can be obtained from Equation~\ref{eq:C} as:
\begin{align}   
    \mathcal{C}(\vec{x},\vec{y}) = \mathcal{J}(\vec{x},\vec{y}) \, \mathcal{I}(\vec{x},\vec{y}) = \frac{2.2}{10.5} \frac{2.2}{6.2}
    = \frac{4.84}{65.1} \approx 0.074  \nonumber
\end{align}

The resulting value is particularly small, indicating that the two vectors are mostly distinct one another, which is indeed the case.

Given a dataset with $N$ elements characterized by $M$ respective properties, it is possible to obtain the respective \emph{coincidence similarity network}~\cite{CostaCCompl} by representing each data element as a respective node, while the pairwise interconnections have weights corresponding to the respective coincidence similarity index between that pair of data elements.

This type of data representation has two especially interesting properties: (i) they links directly reflect the \emph{similarity} between pairs of data elements; and (ii) the weights of the links are intrinsically normalized, with values in the interval $[0,1]$. Therefore, groups of data elements will tend to appear as connected components or modules (communities) in the obtained similarity networks, so that connected component or community detection methods can be employed to identify the respective possible groups or clusters.

\begin{figure}[h!]
  \centering
     \includegraphics[width=.45 \textwidth]{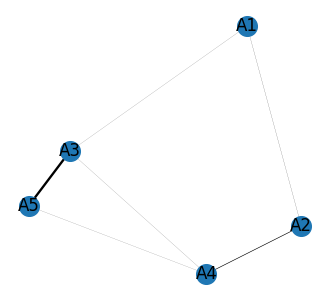}  \hspace{1cm}
     \includegraphics[width=.45 \textwidth]{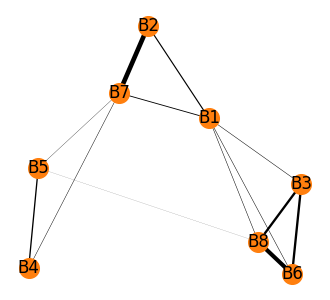}  \\
     (a) \hspace{6cm} (b)
   \caption{The coincidence similarity networks obtained respectively to the representations of the same bipartite presented in Figs~\ref{fig:bipartite_D}(a) and ~\ref{fig:bipartite_D}(b). Tables~\ref{tab:bipartiteA} and~\ref{tab:bipartiteB} were used respectively, with $8$ and $5$ associated features. The width of each link is proportional to the respective coincidence similarity between pairs of nodes, indicating that more similar pairs of nodes are more strongly interconnected. The similarity between the interconnection patterns exhibited by each group is more directly indicated in the depicted coincidence networks than in the unconstrained network shown in Fig.~\ref{fig:bipartite_net} and provides more detailed information about the interconnections than the projections of the original bipartite network shown in Fig.~\ref{fig:Proj}.}\label{fig:coinc_net_AB}
\end{figure}

Figure~\ref{fig:coinc_net_AB} depicts the coincidence similarity networks derived from the non-directed bipartite network in Figure~\ref{fig:bipartite} by considering its direct (a) and reverse (b) representations illustrated respectively in Figs.~\ref{fig:bipartite_D}(a) and ~\ref{fig:bipartite_D}(b).

It is interesting to observe that markedly distinct coincidence similarity networks have been obtained respectively to the direct and reverse representations of the \emph{same} non-directed bipartite network in Figure~\ref{fig:bipartite_D}.  When compared to the projections of the original bipartite network shown in Figure~\ref{fig:Proj}, the obtained coincidence similarity networks provide more detailed information about the respective interconnectivity as a consequence not only of taking into account the weights of the original bipartite links, but also because of the enhanced selectivity provided by the coincidence similarity index.

When comparing the two obtained coincidence similarity networks with the representation of the original non-directed network irrespectively to the types of its nodes shown in Figure~\ref{fig:bipartite_net}, it is interesting to observe that, though this representation includes all types of nodes, it does not provide a direct indication of the relationships between the elements of each of the two types. For instance, the fact that elements $A3$ and $A5$ are strongly similar, and therefore interrelated, is immediately indicated in the coincidence similarity in Figure~\ref{fig:coinc_net_AB}(a), while being much harder to infer indirectly from the representation in Figure~\ref{fig:bipartite_net}.

\begin{figure}[h!]
  \centering
     \includegraphics[width=.45 \textwidth]{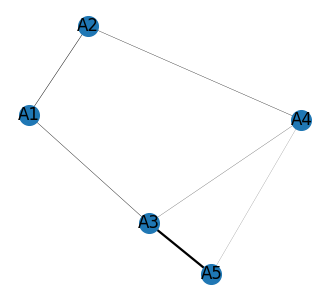}  \hspace{1cm}
     \includegraphics[width=.45 \textwidth]{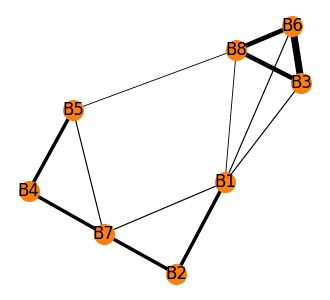}  \\
     (a) \hspace{6cm} (b)
   \caption{The coincidence similarity networks of the original bipartite network in Fig.~\ref{fig:bipartite} obtained while considering all weights equal to 1 (unweighted version).}\label{fig:coinc_netW1_AB}
\end{figure}

It is of particular interest to compare the obtained coincidence similarity networks with the projections of the original bipartite obtained \emph{while taking all weights being equal to 1}, which are shown in Figure~\ref{fig:coinc_netW1_AB}.

Though the respective projected and coincidence similarity networks tend to be similar, there are important differences. First, as indicated when comparing the $A$ nodes networks, the coincidence similarity network tends to present more detailed information about the interconnections. Second, as can be inferred while comparing the $B$ nodes networks, a markedly distinct interconnectivity can be verified. This is a consequence not only of the enhanced selectivity of the coincidence similarity network but also of the fact that the similarity between nodes takes into account all their links to other nodes of other type, while only the singly shared nodes of the other type determine the interconnections in the projection approach.

\subsection{Generating Synthetic Bipartite Networks}

While real-world data is always of interest, it is often the case that synthetically generated data can provide a more objective and systematic evaluation of performance as a consequence of the properties of the data being fully specified by a respective model. The present work studies the performance of the described coincidence similarity approach to identifying groups in bipartite networks by considering data generated synthetically as described in the following.

First, the number of groups ($N_g$) of each network is specified, as well as the number of respective nodes of type $A$ and $B$ in each group ($N_A$ and $N_B$, respectively). This implies in a total of $N_g \, N_A$ nodes of type $A$ and $N_g \, N_B$ nodes of type $B$.  Therefore, for simplicity's sake, all groups have the same number of nodes of type $A$ and $B$, implying that each node of type $A$ in each group is connected to all nodes of type $B$ in that same group.

The initial interconnection pattern described above leads to $N_g$ completely separated groups (connected components in the original bipartite network). In order to incorporate overlap between groups, which is almost invariably observed in real-world bipartite networks, the initial interconnections are scrambled with probability $p$. This scrambling procedure is performed by exchanging the position of each link between nodes of different types with uniform probability.

\section{A Preliminary Case-Example}\label{sec:example}

The identification of groups in non-directed bipartite networks by using the coincidence similarity approach has a particularly important aspect regarding the possibly distinct results when considering the direct and reverse respective directed bipartite representations, as it has already been observed respectively to the Figures~\ref{fig:coinc_net_AB}(a) and ~\ref{fig:coinc_net_AB}(b). Therefore, it is important to consider this issue in more detail before proceeding to a more systematic study of the performance of the similarity approach, which is described as follows respectively to a simple case-example.

\begin{figure}[h!]
  \centering
     \includegraphics[width=.9 \textwidth]{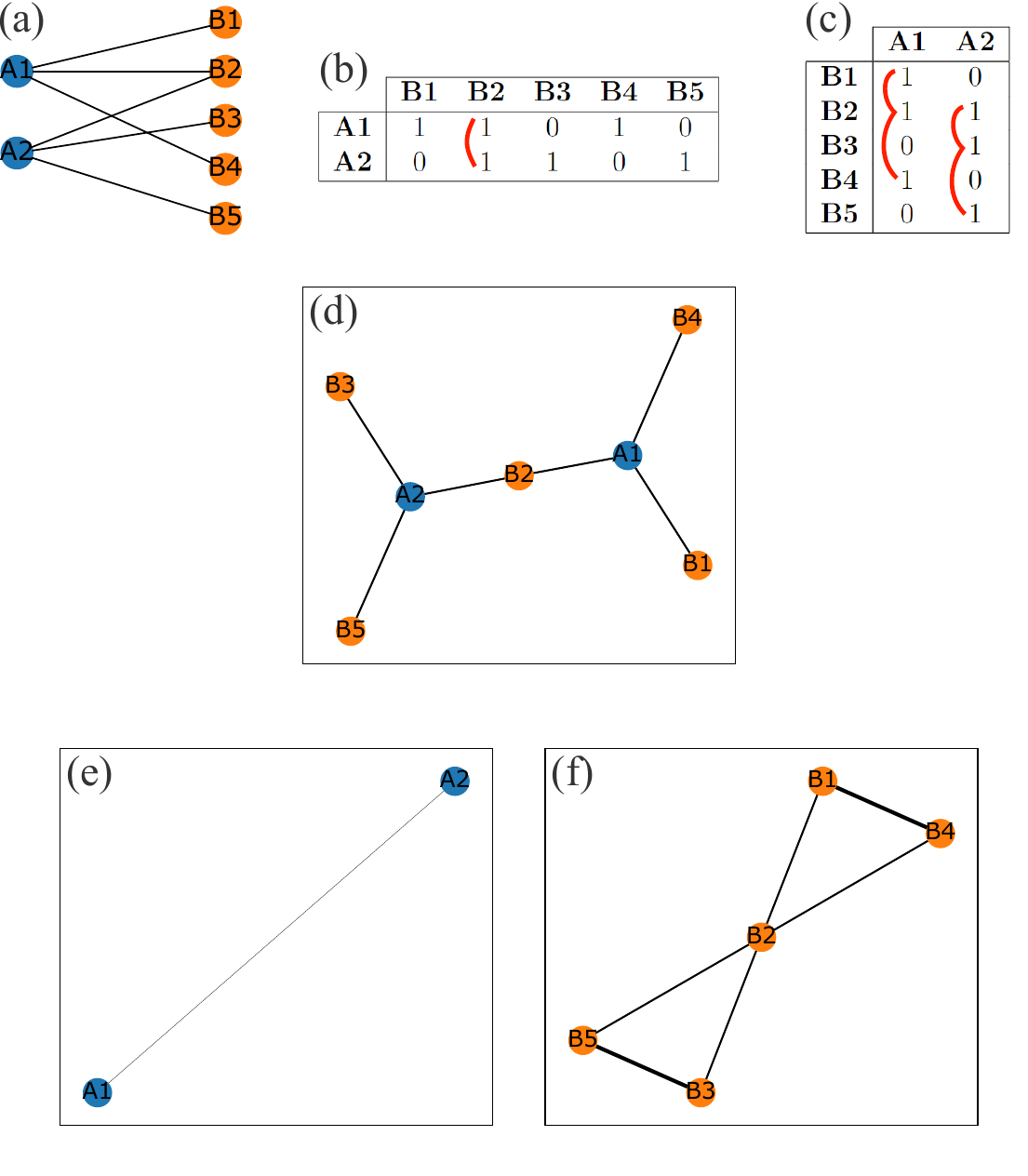}
   \caption{(a) A simple bipartite networks incorporating two groups ($A1$ and $A2$), each having a single element. These two nodes can be considered as two distinct groups because they present mostly distinct interconnections with the elements of type $B$, except for the two links involving the node $B2$. The interconnection tables of the respective direct and reverse representations are presented in (b) and (c), respectively. The overall network shown in an unconstrained manner is presented in (d). The coincidence networks obtained for the direct and reverse representations are depicted in (e) and (f), respectively.}\label{fig:example}
\end{figure}

Figure~\ref{fig:example}(a) illustrates a simple non-directed bipartite network with weights being restricted to the values 0 and 1. This bipartite network can be understood as presenting two groups of a single element which ($A1$ and $A2$), characterized by mostly distinct respective patterns of interconnections with the nodes of type $B$ (related to groups $B1,B2,B4$ and $B2,B3,B5$). Observe that a single interconnection is shared between these two groups, namely that involving the node $B2$. This figure also presents the direct (b) and reverse (c) interconnections matrix, the unconstrained visualization of the original bipartite network (d), as well as the direct (e) and reverse (f) coincidence networks.

When comparing the networks obtained respective to the original bipartite network (d) and the two coincidence networks in (e) and (f), it can be immediately observed that the direct coincidence network shown in (e) provides the best separation between the two groups, with the two nodes being linked by the thinnest link in all presented networks. That is because there are substantially more columns (features) in Table (b) than in Table (c), allowing more possibilities for having distinct patterns of interconnections between the two groups in the latter case.

The enhanced separation between the groups allowed by the direct representation is related to the fact that only the two entries identified by the red arc in Figure~\ref{fig:example}(b) present identical feature values for the two distinct groups. At the same time, there are several identical features shared by between the two groups, identified by red arcs in Figure~\ref{fig:example}(c).

It is indeed known from the area of pattern recognition (e.g.~\cite{duda1973pattern,costa2000shape}) that the separation between groups tends to be enhanced by the incorporation of additional (but not excessive) properties or features of each of the elements giving rise to the groups. Indeed, in the case of the specific example above, it is impossible to have completely distinct features of five elements using only two binary properties (the maximum number of distinct patterns, in this case, would be only $2^2=4$).

If follows from the above considerations that enhanced separation between the existing groups tends to be obtained from the directed representation having a larger number of features (columns).

\section{Experimental Results}\label{sec:results}

In this section, the experiments performed in order to study the coincidence similarity approach to identifying groups in bipartite networks are described and discussed. For the sake of comprehensiveness, we resource to synthetic bipartite networks whose groups are preliminarily known.

We consider synthetic bipartite networks involving $N_g=3, 5,$ and $7$ groups, each with $N_A=5$ elements of type $A$ and $N_B=10$ elements of type $B$. The interconnections are rewired (scrambled) with probability $p$.

\begin{figure}[h!]
  \centering
     \includegraphics[width=.99 \textwidth]{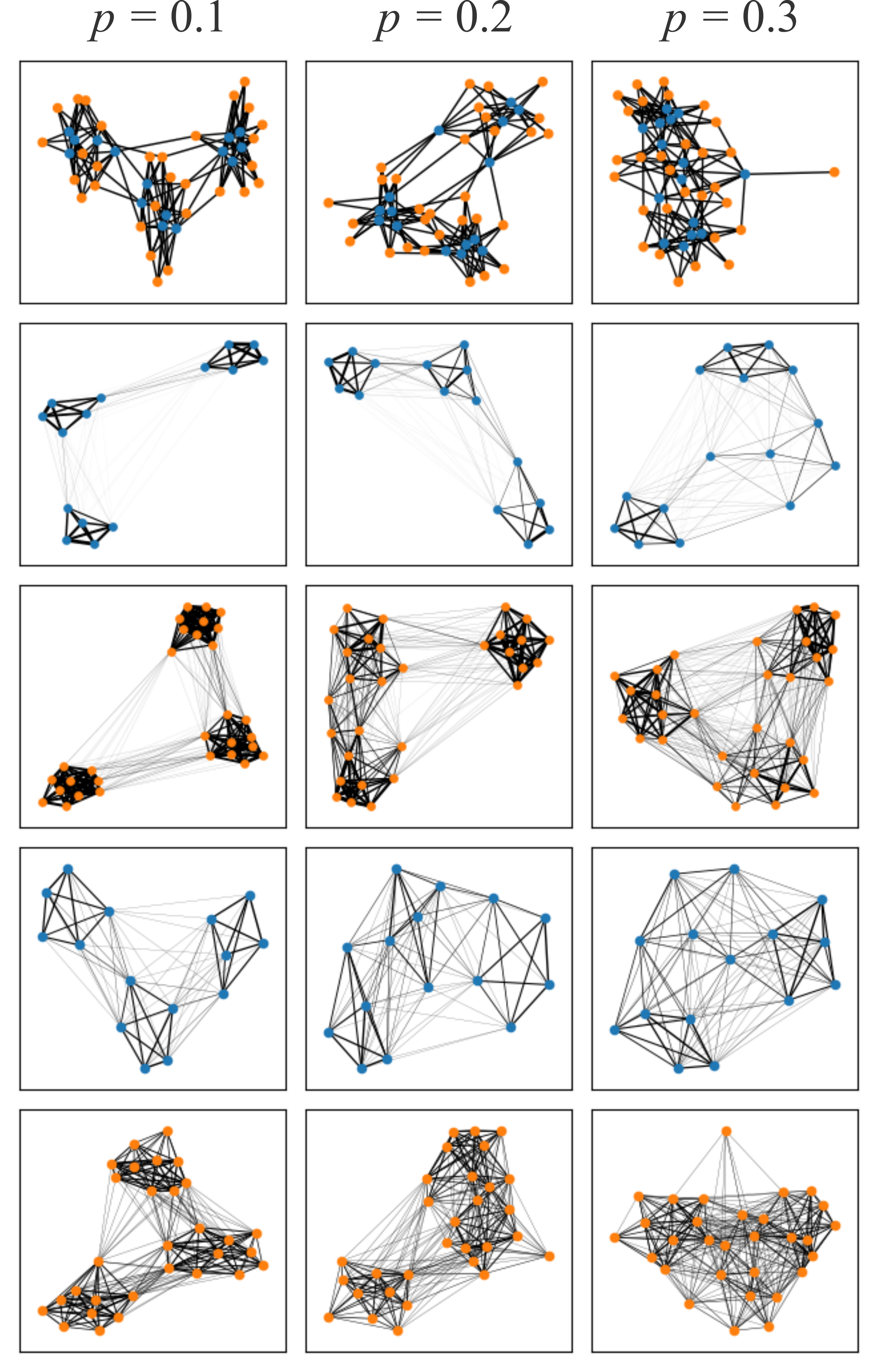}
   \caption{Unconstrained (first row), direct coincidence similarity (second row), reverse coincidence similarity (third row), projection onto $A$ nodes (fourth row), and projection onto $B$ nodes (fifth row) networks considering $N_g=3$ groups and rewiring levels $p=0.1, 0.2,$ and $0.3$. The best separations have been obtained for the direct coincidence similarity networks in all cases, followed by the reverse coincidence similarity structures.}\label{fig:comparison_3}
\end{figure}

Figure~\ref{fig:comparison_3} presents the unconstrained (first row), direct coincidence similarity (second row), reverse coincidence similarity (third row), projection onto $A$ nodes (fourth row), and projection onto $B$ nodes (fifth row) networks obtained for a bipartite network with 3 groups considering three rewiring levels $p=0.1, 0.2,$ and $0.3$. It should be observed that a rewiring by $p=0.3$ is quite intense, implying a respective perturbation of almost one third of the links in the original bipartite network.

The networks obtained by using the coincidence similarity approach resulted with well-separated groups in all cases, even in the presence of $30\%$ rewiring of the original bipartite network. At the same time, as could be expected in the light of the discussion in Section~\ref{sec:example}, the greatest separation has been obtained for the case involving the largest number of features (columns) which, in all cases, corresponds to the direct coincidence similarity networks shown in the second column.

\begin{figure}[h!]
  \centering
     \includegraphics[width=.99 \textwidth]{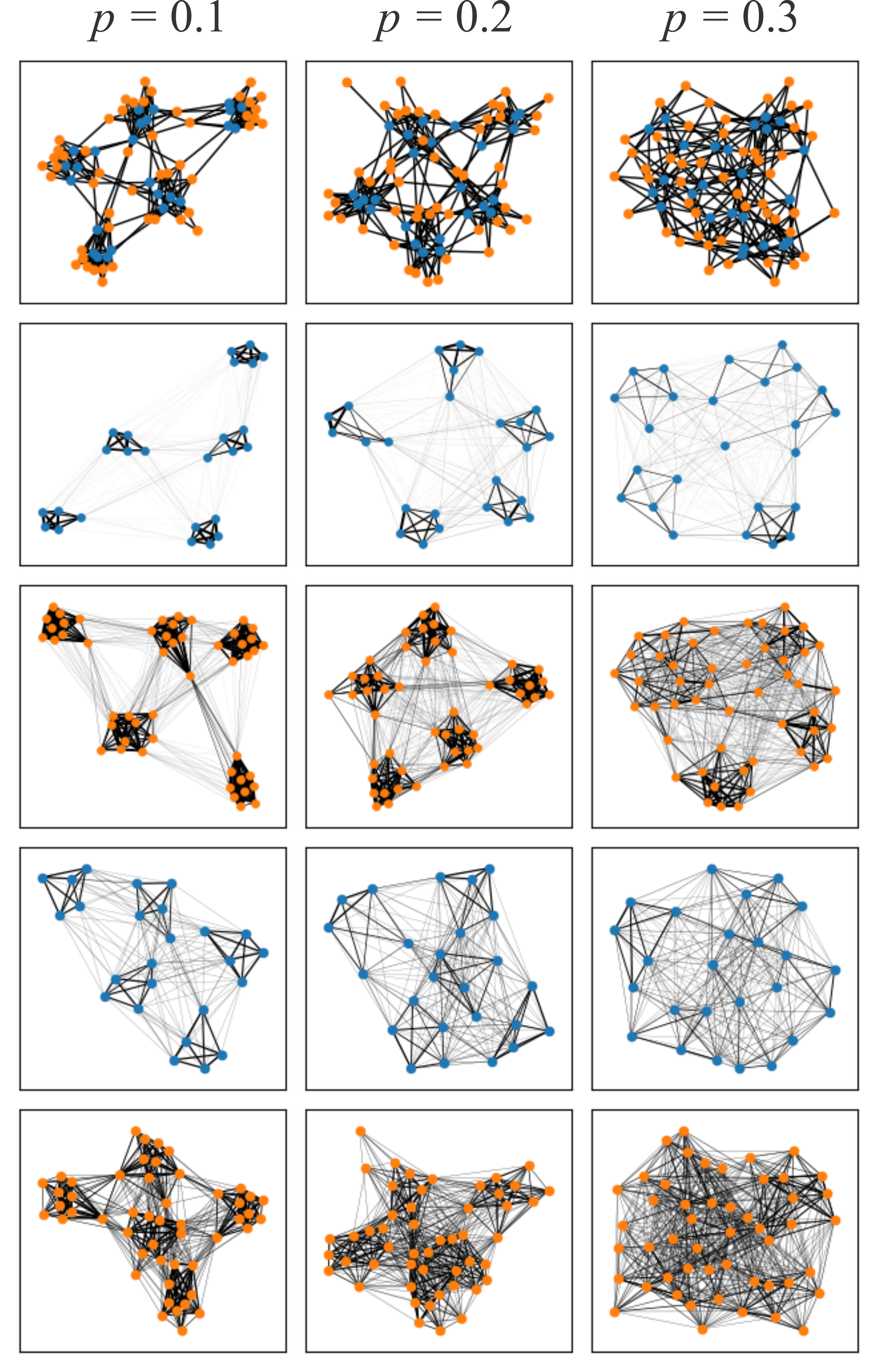}
   \caption{Unconstrained (first row), direct coincidence similarity (second row), reverse coincidence similarity (third row), projection onto $A$ nodes (fourth row), and projection onto $B$ nodes (fifth row) networks considering $N_g=5$ groups and rewiring levels $p=0.1, 0.2,$ and $0.3$. The best separations have been obtained for the direct coincidence similarity networks in all cases, followed by the reverse coincidence similarity structures.}\label{fig:comparison_5}
\end{figure}

\begin{figure}[h!]
  \centering
     \includegraphics[width=.99 \textwidth]{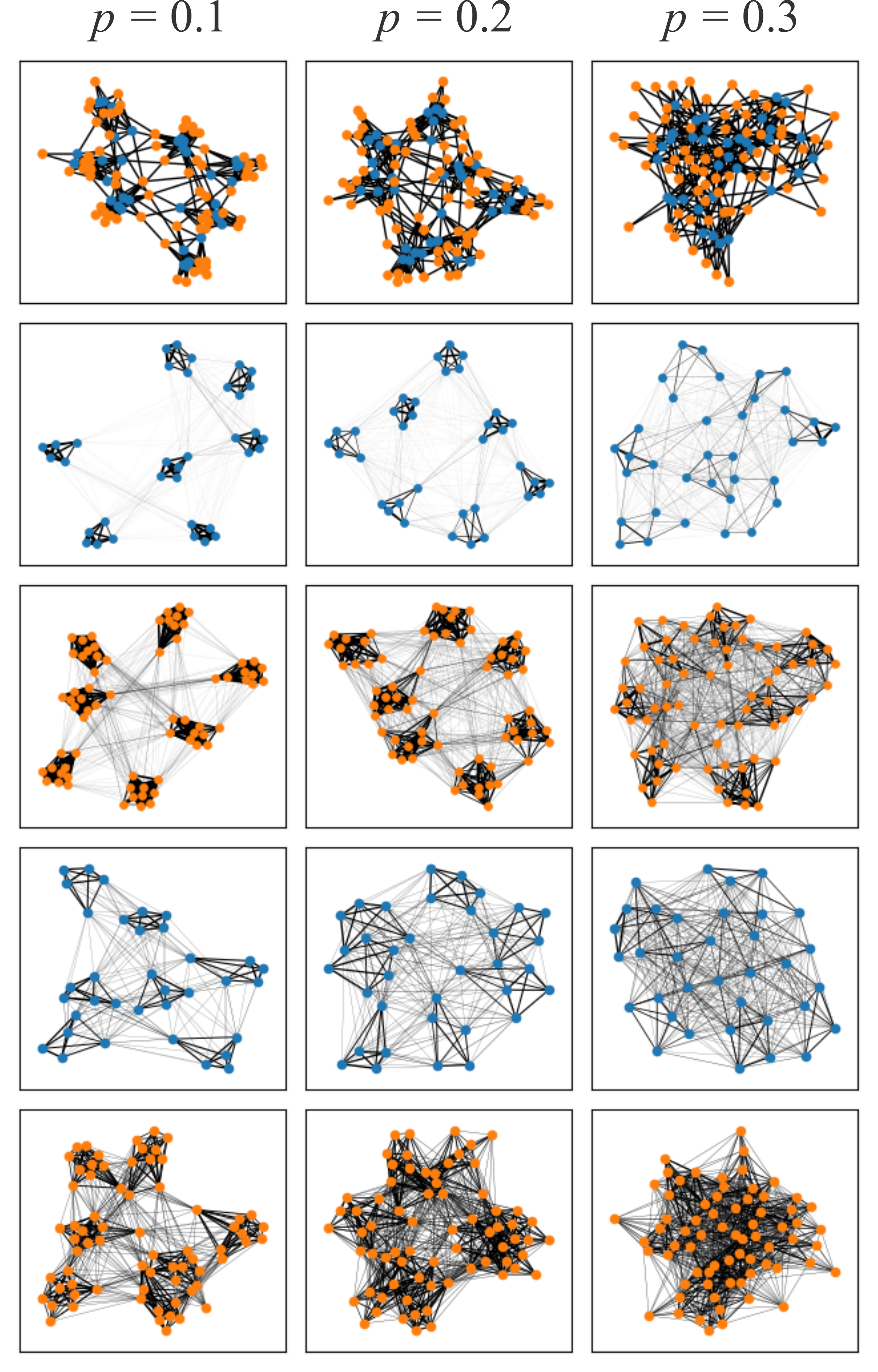}
   \caption{Unconstrained (first row), direct coincidence similarity (second row), reverse coincidence similarity (third row), projection onto $A$ nodes (fourth row), and projection onto $B$ nodes (fifth row) networks considering $N_g=7$ groups and rewiring levels $p=0.1, 0.2,$ and $0.3$. The best separations have been obtained for the direct coincidence similarity networks in all cases, followed by the reverse coincidence similarity structures.}\label{fig:comparison_7}
\end{figure}

The results obtained for $N_g=5$ and $N_g=7$ groups are depicted in Figures~\ref{fig:comparison_5} and~\ref{fig:comparison_7}, respectively. As with the case $N_g=3$, the best results were obtained for the direct coincidence similarity networks (second column in the figure).

In most cases, the original groups can be recovered by performing a simple thresholding on the interconnecting weights. Those cases involving more intense group overlap can be addressed by applying some community detection methodology (e.g.~\cite{newman2004detecting,rosvall2009map}).

\section{Quantitative Performance Analysis}

Though the results presented in Section~\ref{sec:results} illustrate in a well-defined manner the potential of the coincidence similarity approach to detecting groups in bipartite networks, it is still interesting to consider respective quantitative performance analysis.

One of the advantages of adopting synthetic data is that the original groups and respective properties are completely known, thus allowing a fully objective quantification of the detection errors to be considered in the performance evaluation. More specifically, here we consider the percentage of errors while estimating the links between the original groups ($\epsilon_b$) and the percentage of errors respective to the links within those groups ($\epsilon_w$).

The two errors above can be obtained by considering the original groups in the synthetic data and the results of the group identification by the coincidence similarity methodology. More specifically, the resulting coincidence similarity network is thresholded by a value $T$, and the links are understood to indicate relationships within all nodes of each of the existing groups. Therefore, ideally, the groups should result as completely separated graphs and all nodes within each of these graphs should be completely interconnected, a situation which is characterized by $\epsilon_b=0\%$ and $\epsilon_w=0\%$.

Two errors are considered in the present work, instead of a single figure of merit, as this allows more objective and complete quantification of the obtained performances. Using a single error would have the disadvantage of mixing the results concerning the interconnectivity between and within the original groups.

Because of the enhanced potential of the coincidence similarity approach to group identification, several situations in practice could be addressed simply in terms of the above adopted thresholding approach. However, community finding methodologies may be required in cases where the groups are not well separated.

\begin{figure}[h!]
  \centering
     \includegraphics[width=.32 \textwidth]{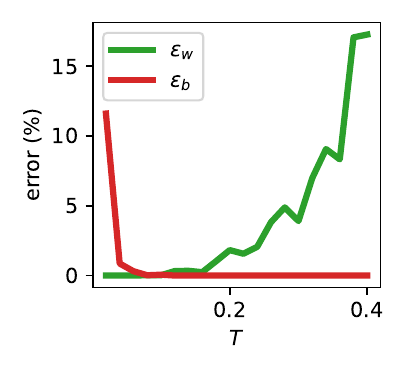}
     \includegraphics[width=.32 \textwidth]{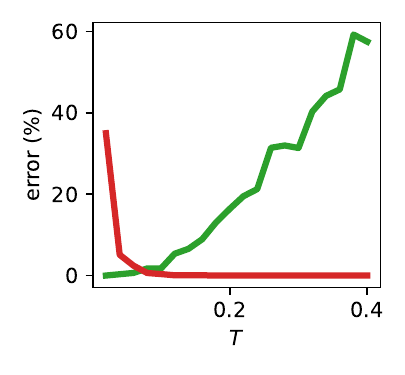}
     \includegraphics[width=.32 \textwidth]{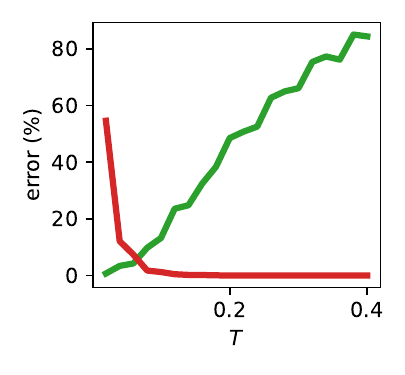} \\
     \hspace{.5cm} (a) \hspace{3.7cm} (b) \hspace{3.7cm} (c) \\ \vspace{0.7cm}
     \includegraphics[width=.32 \textwidth]{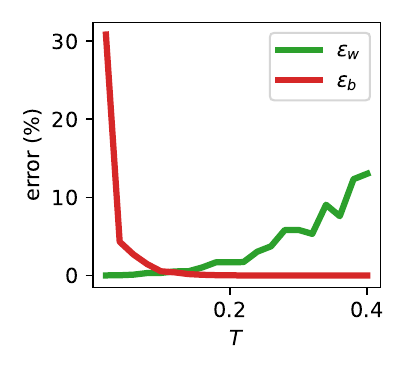}
     \includegraphics[width=.32 \textwidth]{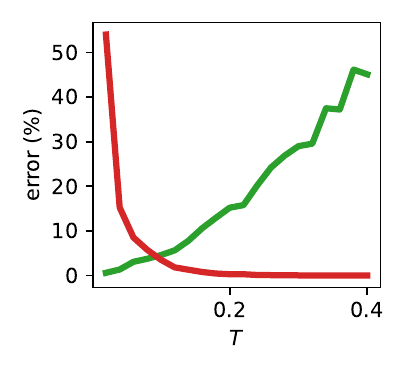}
     \includegraphics[width=.32 \textwidth]{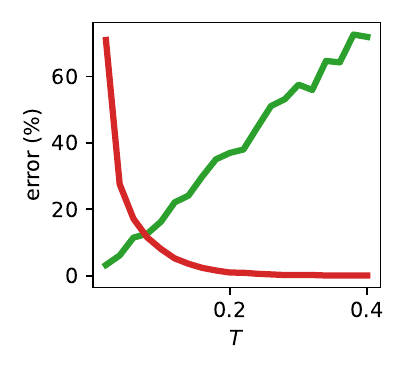} \\
     \hspace{.5cm} (d) \hspace{3.7cm} (e) \hspace{3.7cm} (f)
   \caption{The errors $\epsilon_b$ and $\epsilon_w$ in terms of the threshold value ($T$) obtained for 100 realizations of the group identification case addressed in Section~\ref{sec:results} respectively to $N_g=5$, $N_A=5$, $N_B=10$ and $p=0.1$, $p=0.2$, and $p=0.3$. The errors obtained by considering the \emph{direct} and \emph{reverse} coincidence similarity networks are shown in the first (a-c) and second (d-f) rows, respectively. The $\epsilon_b$ error decreases with the threshold value, while larger values of the $\epsilon_w$ error are respectively obtained. As could be expected, both errors increase with the rewiring probability $p$. Larger errors $\epsilon_b$ have been obtained for the reverse coincidence similarity network as a consequence of this representation involving fewer number of columns (features).}\label{fig:errorA}
\end{figure}

Figure~\ref{fig:errorA} illustrates the $\epsilon_b$ and $\epsilon_w$ errors respectively to 100 realizations of the situation addressed in Section~\ref{sec:results} respectively to $N_g=5$, $N_A=5$, $N_B=10$ and $p=0.1$, $p=0.2$, and $p=0.3$. The first (a-c) and second (d-f) rows in this figure contains the errors obtained while considering the \emph{direct} and \emph{reverse} coincidence similarity networks, respectively.

The $\epsilon_b$ error decreases with the threshold values in all situations, while the $\epsilon_w$ error increases in that same situation. That is because higher threshold values tend to increase the separation between the groups and to decrease the interconnection between nodes within the same group.

Both types of errors tend to increase steadily with $p$. Of particular interest is the fact that the $\epsilon_b$ error tends to decrease sharply with the threshold in all considered situations, allowing a small value of $T$ to be chosen which also implies that the $\epsilon_w$ error also results small. Indeed, in the case of \emph{direct} coincidence similarity networks, it is possible to choose a threshold value that allows nearly $0\%$ errors in the case of $p=0.1$ and $0.2$. The results obtained for the \emph{reverse} coincidence similarity networks are characterized by relatively larger error $\epsilon_b$, which corroborate the tendency of obtaining less separated groups in case of using a smaller number of features (in the specif cases considered here, this corresponds to the reverse coincidence networks).

In addition, it is interesting to observe that moderate values of $\epsilon_w$ tend to have little impact for group identification, as not every internal link is necessary in order to characterize an isolated group (basically, groups tend to have substantially more inner connections than outer connections). These results are particularly important because it corroborates the effectiveness of the coincidence similarity approach for identifying groups in bipartite networks.

\section{Concluding Remarks}

Bipartite networks constitute a specific, but particularly important type of graphs and complex networks. A good deal of the interest in bipartite networks relates to the identification of respective groups, clusters, or modules of nodes.

Though many previous approaches have resourced to the concept of modularity for identifying groups in unconstrained representations of bipartite networks, the consideration of similarity indices represents a potentially interesting alternative approach that can be used to identify groups in bipartite networks~\cite{costa_biorXiv}. 

In the present work, we expanded and complemented the previous approach described in~\cite{costa_biorXiv} in several ways, including the characterization of non-directed and directed bipartite networks, more systematic consideration of what are clusters in bipartite networks, as well as the description of an approach to obtain a virtually unlimited number of synthetic bipartite networks with a specific number of groups, number of elements per group, and varying types of interconnectivity. The latter has then been used as a resource for inferring the performance of the coincidence similarity approach as a means of identifying groups or clusters of nodes in bipartite networks.

Several interesting results have been described and discussed, including the tendency of obtaining enhanced separation between the groups by choosing the directed representation involving the largest number of columns (features). In addition, the potential of the coincidence similarity approach to identifying groups in bipartite network previously hinted in~\cite{costa_biorXiv}, has been corroborated by experiments involving synthetic bipartite numbers presenting a pre-specified number of groups, number of elements per group, as well as varying types of interconnection patterns. In all considered experimental situations, the coincidence similarity approach led to enhanced separation between the groups when compared to the modularity of the original bipartite network. Another particularly interesting result described in the present work concerns the tendency to obtained enhanced group separation by adopting the coincidence similarity network (direct or reverse representations) having the larger number of columns (features).

The reported concepts, methods, and results pave the way to a number of related future developments, which include but are not limited to the following. First, the performance analysis could be extended to bipartite networks involving groups of distinct sizes. It would also be interesting to revisit results previously obtained by using the modularity of real-world bipartite networks in terms of the coincidence similarity methodology, possibly leading to respective complementation of the analysis. Another prospect of particular interest regards the adaptation of the described methodology to identify groups in multipartite networks. More generally, it would be possible to search for groups in a given subset of nodes of a network respectively to their interconnections with the remainder nodes. Yet another possible work of consideration would be to adapt the proposed methodology to identify modules (communities) in networks potentially involving two types of nodes.

\section*{Acknowledgments}
Alexandre Benatti thanks MCTI PPI-SOFTEX (TIC 13 DOU 01245.010222/2022-44).
Luciano da F. Costa thanks CNPq (grant no.~307085/2018-0) and FAPESP (grants 15/22308-2 and 2022/15304-4).

\bibliography{ref}
\bibliographystyle{unsrt}

\end{document}